\documentclass{PoS}

\usepackage{amsmath, amssymb, amsfonts}

\newcommand{\Phibar}{\ensuremath{\bar{\Phi}}}
\def\dr{{D\!\llap{/}}\,}
\newcommand{\Tr}{\ensuremath{\operatorname{Tr}}}

\def\eq#1{Eq.~(\ref{#1})}
\def\fig#1{Fig.~\ref{#1}}
\def\tab#1{Tab.~\ref{#1}}

\definecolor{bjcol}{rgb}{1,0.205,0.80}

\title{Exploring the Phase Structure and Thermodynamics of QCD}

\ShortTitle{Thermodynamics of QCD}

\author{\speaker{Tina K. Herbst}\\
        Institut f\"ur Theoretische Physik,
	Universit\"at Heidelberg, 
	Philosophenweg 16, D-69120 Heidelberg, Germany\\
        E-mail: \email{t.herbst@thphys.uni-heidelberg.de}}

\author{Mario Mitter\\
	Institut f\"ur Theoretische Physik, Universit\"at Heidelberg, 
	Philosophenweg 16, D-69120 Heidelberg, Germany}

\author{Jan M. Pawlowski\\
	Institut f\"ur Theoretische Physik, Universit\"at Heidelberg, 
	Philosophenweg 16, D-69120 Heidelberg, Germany and
	ExtreMe Matter Institute EMMI, GSI Helmholtzzentrum f\"ur 
	Schwerionenforschung mbH, Planckstra{\ss}e 1, D-64291 Darmstadt,
	Germany}

\author{Bernd-Jochen Schaefer\\
	Institut f\"{u}r Theoretische Physik,
        Universit\"at Giessen,
	Heinrich-Buff-Ring 16, D-35392 Giessen, Germany and
	Institut f\"{u}r Physik, Karl-Franzens-Universit\"{a}t Graz,
	Universit\"atsplatz 5, A-8010 Graz, Austria}

\author{Rainer Stiele\\
	Institut f\"ur Theoretische Physik, Universit\"at Heidelberg, 
	Philosophenweg 16, D-69120 Heidelberg, Germany and
	ExtreMe Matter Institute EMMI, GSI Helmholtzzentrum f\"ur 
	Schwerionenforschung mbH, Planckstra{\ss}e 1, D-64291 Darmstadt,
	Germany}

\abstract{\noindent 
We put forward a Polyakov-loop extended quark meson model, where matter as well
as glue fluctuations are taken into account, cf. \cite{Herbst:2013ufa}. The
latter are included via a Polyakov-loop potential. Usually such a glue potential
is based on Yang-Mills lattice data only. We show that a parametrisation of
unquenching effects as proposed in \cite{Haas:2013qwp}, together with the
inclusion of fluctuations via the functional renormalisation group
\cite{Herbst:2013ail, Mitter:2013fxa}, accounts for the relevant dynamics.
This is demonstrated by a comparison of order parameters and
thermodynamic observables to recent lattice results at vanishing chemical
potential, where we find very good agreement.}

\FullConference{QCD-TNT-III-From quarks and gluons to hadronic matter: A bridge
		too far?,\\
		2-6 September, 2013\\
		European Centre for Theoretical Studies in Nuclear Physics and
		Related Areas (ECT*), Villazzano, Trento (Italy)}

\begin{document}

\section{Introduction}
The understanding of the phase structure and thermodynamics of Quantum
Chromodynamics (QCD) has been a major aim of theoretical as well as experimental
high-energy physics over the last years. Running and planned experiments at
CERN, RHIC, NICA and FAIR will provide us with important insights into the
behaviour of strongly interacting matter under extreme conditions, such as high
temperatures and densities.
For a comprehensive theoretical understanding of these systems, it is necessary
to employ non-perturbative techniques. These allow to study the physics of,
e.g., the chiral and deconfinement phase transitions in QCD. Functional methods,
such as the Functional Renormalisation Group (FRG) provide one such tool. The
recent years have seen a lot of progress in this area: for example it has been
demonstrated that the FRG is a suitable tool to study QCD at finite temperature
as well as at finite chemical potentials \cite{Braun:2007bx, Braun:2009gm,
Fister:2013bh, Fischer:2013eca}.

While first-principles studies of QCD are possible within the FRG framework by
now, they still pose a formidable task. Thus, it has proven fruitful to
employ functional techniques also to low-energy effective models, which share
some aspects with the full theory while being more tractable.
Furthermore, it has been demonstrated that these effective models can be related
to QCD in a systematic fashion, see e.g. \cite{Haas:2013qwp, Herbst:2013ail, 
Pawlowski:2010ht} for a detailed discussion. 
In the present work we use the Polyakov--quark-meson (PQM) model, as one
representative of Polyakov-loop enhanced chiral models, see e.g.
\cite{Fukushima:2003fw, Megias:2004hj, Ratti:2005jh, Mukherjee:2006hq,
Schaefer:2007pw}, which is known to describe the chiral properties of QCD very
well. Furthermore, the coupling to the Polyakov-loop provides a statistical
implementation of confinement. Unquenching effects are included as in
\cite{Herbst:2013ufa, Haas:2013qwp}, see also our discussion in
Sec.~\ref{sec:unquenching}.

A different first-principles approach to QCD are lattice simulations.
These have shed light on the phase structure and bulk thermodynamics,
especially at vanishing chemical potential, where the infamous sign problem is
absent. In this region, results at physical masses \cite{Bazavov:2011nk,
Bazavov:2012bp} and in the continuum limit are available by now \cite{
Borsanyi:2010bp, Borsanyi:2010cj, Borsanyi:2013bia}. These results can be used
to scrutinise results from other approaches, allowing us to determine whether
all relevant physical effects have been included, for a more detailed 
discussion see \cite{Herbst:2013ufa}.

\section{QCD at Low Energies: The Polyakov-Quark-Meson Truncation}
We briefly recapitulate the Polyakov--quark-meson (PQM) 
model~\cite{Schaefer:2007pw} as a low-energy effective theory for QCD. For a
more detailed discussion, the reader is referred to e.g.~\cite{Herbst:2013ail,
Mitter:2013fxa, Schaefer:2008hk, Herbst:2010rf}.  
The Euclidean Lagrangian of this model, including a flavour-blind chemical
potential $\mu$, reads
\begin{eqnarray} \label{eq:Lpqm}
  \mathcal L_{\rm PQM} & = & \bar{q} \left(\dr + h\,T^a (\sigma_a 
  + i \gamma_5 \pi_a ) + \mu\gamma_0\right) q 
  +\ \mathcal L_{m} +\ V_{\rm glue}(\Phi,\Phibar;t) \,.
\end{eqnarray}
Here, the chiral sector is given by the well-known quark-meson model,
see e.g. \cite{Schaefer:2004en} and references therein. It is supplemented by a
mesonic Lagrangian~\cite{Mitter:2013fxa, Schaefer:2008hk}
\begin{eqnarray}\label{eq:Lmeson}
  \mathcal L_m & = & \Tr(\partial_\mu \Sigma\partial_\mu\Sigma^\dagger) +
  U(\rho_1,\tilde\rho_2) + c\,\xi 
   -\ \Tr\left[C(\Sigma+\Sigma^\dagger)\right]\,,
\end{eqnarray}
where $\Sigma$ is a complex matrix containing the mesonic fields: $\Sigma =
\Sigma_a T^a = (\sigma_a + i \pi_a)T^a\,,$ with $\sigma_a$ denoting the scalar
and $\pi_a$ the pseudo-scalar meson nonets. Furthermore, $T^a=\lambda^a/2\,$
denote the Hermitian generators of the flavour $U(3)$ symmetry, defined via
the Gell-Mann matrices.
A flavour-blind Yukawa coupling $h$ couples the quark fields to the mesonic
sector. In the following we will consider the $(2+1)$-flavour case, hence
assuming isospin symmetry in the light sector. 

The meson potential $U$ can be expressed via the chiral invariants $\rho_i~=~
\Tr\left[(\Sigma\Sigma^\dagger)^i\right]$, $i=1,\,\dots,\,N_f$ 
\cite{Jungnickel:1995fp}. In the ($2+1$)-flavour approximation considered here,
we restrict ourselves to the two perturbatively renormalisable invariants $
\rho_1$ and $\tilde\rho_2 = \rho_2 - \frac{1}{3}\rho_1^2$. 
The chiral $U_A(1)$ anomaly is included with the help of the 't\,Hooft
determinant $ \xi = \det(\Sigma) +
\det(\Sigma^\dagger)$ \cite{Hooft:1976fv, Hooft:1976up}. The strength of
its coupling, $c$, determines the mass splitting between the
$\eta$, $\eta'$ and pions, see e.g. \cite{Mitter:2013fxa, Schaefer:2008hk, 
Schaefer:2013isa} for a detailed discussion.

The gauge fields, represented in terms of Polyakov-loop variables, are coupled 
to the matter sector via the covariant derivative $\dr(\Phi) = \gamma_\mu
\partial_\mu-i\,g \gamma_0 A_0(\Phi)$.
Integrating out the gluonic degrees of freedom furthermore results in a
potential for the Polyakov loops, $V_{\rm glue}(\Phi(A_0),\,\Phibar(A_0))$,
which is discussed in more detail in the following.

\section{Unquenching}\label{sec:unquenching}
Consider the effective action of full QCD, which can be written as
\begin{equation}\label{eq:QCDaction}
  \Gamma_k = \beta \mathcal V V[A_0] +\Delta\Gamma_k[\bar A_0, \phi]\,,
\end{equation}
where $\mathcal V$ is the spatial volume and $\beta=1/T$ the inverse 
temperature. In \eq{eq:QCDaction}, the first term denotes the QCD glue 
potential, encoding the ghost-gluon dynamics in the presence of matter fields. 
The second term contains the matter contribution coupled to a background gluon 
field $\bar A_0$. For brevity, the remaining field content is collected in
$\phi$. This part is well-described in terms of low-energy chiral models, such
as the PQM model discussed above.

It is important to notice that the QCD glue potential is different from its
pure Yang-Mills counterpart: unquenching effects of the matter fields
modify the QCD glue potential. In particular, they lower the deconfinement
temperature, see also the discussion in \cite{Herbst:2013ufa, Haas:2013qwp,
Pawlowski:2010ht, Schaefer:2007pw, Herbst:2010rf}. Within effective model
studies, different ans\"atze for this potential are used, which usually are
fitted to Yang-Mills lattice data. Obviously, this procedure does not account
for the unquenching effects.

By now, first principles FRG calculations of the glue potential in pure
Yang-Mills theory as well as 2-flavour QCD in the chiral limit are available
\cite{Braun:2007bx, Braun:2009gm, Fister:2013bh, Braun:2010cy}. These have
revealed striking similarities between the two potentials: apart from a linear
rescaling and shift in the critical temperature, their overall shape is very
similar. This insight has been used in \cite{Haas:2013qwp} to derive a simple
relation between the reduced temperatures, $t_{\rm YM, glue} = ({T - T_{\rm YM,
glue}^{\rm cr}})/{T_{\rm YM, glue}^{\rm cr}}$, of the two theories 
\begin{equation}\label{eq:tYMglue}
  t_{\rm YM }(t_{\rm glue}) \approx 0.57\, t_{\rm glue}\,.
\end{equation}

In practice, \eq{eq:tYMglue} allows  to keep the well-known parametrisations of
the Polyakov-loop potential, as proposed e.g. in \cite{ Ratti:2005jh,  
Pisarski:2000eq, Roessner:2006xn, Fukushima:2008wg, Lo:2013hla}, and include
unquenching effects by a simple rescaling of the reduced temperature. In
\cite{Haas:2013qwp} this approach has been used within a mean-field analysis of
the PQM model. Already there it was found that the inclusion of the unquenching
effects in this simple manner yields results that are very close to the lattice
ones. Here, we report on the next step that consists of including quantum and
thermal fluctuations via the FRG. A more detailed discussion is provided in 
\cite{Herbst:2013ufa}.
Specifically, we show results for a polynomial version of the Polyakov-loop
potential, introduced in~\cite{Ratti:2005jh, Pisarski:2000eq}
\begin{eqnarray} \label{eq:upoly}
  \frac{\mathcal U_{\rm poly}(\Phi,\Phibar;t)}{T^{4}} & = &
  -\frac{b_2(t)}{2}\Phi\Phibar 
   -\ \frac{b_3}{6}\left(\Phi^{3}+\Phibar^{3}\right) +\ \frac{b_4}{4}
  \left(\Phi\Phibar\right)^{2}\,,
\end{eqnarray}
with the (Yang-Mills based) parameters as given in~\cite{Ratti:2005jh}.
The remaining open parameter in \eq{eq:tYMglue}, $T_{\rm cr}^{\rm glue}$, is
fixed by a comparison to the lattice result for the pressure, resulting in
$T_{\rm cr}^{\rm glue}=210$~MeV for our FRG computation. A discussion
of the parameter dependence of our results can be found in
\cite{Herbst:2013ufa}.

\section{Including Fluctuations}
As mentioned above, we include thermal as well as quantum fluctuations with the
FRG. It has been shown previously, see e.g.~\cite{Schaefer:2004en}, that this 
is crucial to achieve a realistic description of the phase transition, which is 
too steep in, e.g., the standard mean-field approximation.

In \cite{Herbst:2013ufa} we have put forward the flow equation for the
$(2+1)$-flavour PQM model in the lowest order of a derivative expansion
\begin{eqnarray}\label{eq:PQM2+1flow}
   \partial_t \Omega_k & = & \frac{k^5}{12\pi^2}
  \left\lbrace \sum_{i=1}^{2N_f^2}\frac{1}{E_i}\coth\left(\frac{E_i}{2T}\right) 
  -\frac{8N_c}{E_{l}}\left[1 - N_{l}( T,\mu;\Phi,\Phibar)
    -  N_{\bar l} ( T,\mu; \Phi,\Phibar)\right] \right.\nonumber\\[1ex]
  && \quad \left. -\frac{4N_c}{E_{s}}\left[1 - N_{s}( T,\mu; \Phi,\Phibar)
    - N_{\bar s} ( T,\mu; \Phi,\Phibar)\right] \right\rbrace\,.
\end{eqnarray}
Here, the Polyakov-loop enhanced quark/anti-quark occupation numbers are given
by
\begin{eqnarray}
  N_{q}(T,\mu;\Phi,\Phibar) & = & \dfrac{1+2\Phibar e^{(E_q-\mu)/T}+\Phi
    e^{2(E_q-\mu)/T}}{1+3\Phibar e^{(E_q-\mu)/T}+3\Phi e^{2(E_q-\mu)/T} +
    e^{3(E_q-\mu)/T}}\,,
\end{eqnarray}
and $N_{\bar q}(T,\mu;\Phi,\Phibar) \equiv N_q(T,-\mu;\Phibar,\Phi)$
for $q=l,s$. Furthermore, the quasi-particle energies of the quarks and mesons
are given by $E_j = \sqrt{k^2+m_j^2}$, $j~\in~\lbrace l,\, s,\, i \rbrace$ with
$i \in \lbrace \sigma,\, \vec{a_0},\, \vec\kappa,\, f_0,\, \vec\pi,\, \vec K,\,
\eta,\, \eta' \rbrace$.

For convenience, we use a non-strange--strange basis, where the light and
strange quark masses can be expressed in terms of the non-strange, $\sigma_x$,
and strange, $\sigma_y$, condensates and the Yukawa coupling as $ m_l =
h\frac{\sigma_x}{2}\,, m_s = h\frac{\sigma_y}{\sqrt{2}}$\,.
The meson masses, on the other hand, are defined via eigenvalues of the Hessian
of the potential
\begin{equation}\label{eq:mesonmasses}
  \lbrace m_j^2\rbrace = \text{eig}\left\lbrace H_\Sigma \left(U(\rho_1,
  \tilde\rho_2) + c\xi\right) \right\rbrace\,.
\end{equation}
Equation (\ref{eq:PQM2+1flow}) encodes the scale dependence
of the mesonic couplings via the scale-dependent effective potential
$\Omega_k$\,, while the running of other interactions containing quarks is
neglected.
The flow equation can be solved once the initial potential is fixed at an
ultraviolet scale $\Lambda=1$~GeV, where we expect our truncation to be a
reasonable description of QCD, cf. our discussion above. For the details
of our numerical method and the used parameters we refer the reader to
\cite{Herbst:2013ufa, Mitter:2013fxa, Strodthoff:2011tz}.

Evolving the flow equation from $\Lambda$ to the infrared, $k\to0$,
fluctuations with momenta smaller than $\Lambda$ are included. However, this
also implies that high modes with momenta $k>\Lambda$ are neglected. In
particular, this translates into a restriction of the tractable temperature
range, $\Lambda\gtrsim2\pi T\,.$ Above this temperature, thermal fluctuations
become important also at scales above the cutoff.
Therefore, the initial potential $\Omega_\Lambda$ is not fully independent of
temperature, which becomes quantitatively important in the region  $2\pi
T\gtrsim \Lambda$. 
In \cite{Herbst:2013ufa} we have argued that the fermionic contributions are
dominant over the bosonic ones in this region. Moreover, we have shown there,
how these fermionic temperature fluctuations can be included in the initial
action $\Omega_\Lambda$ at the cutoff. Our argument relies on the fact that the
temperature dependence of the initial potential $\Omega_\Lambda$ is also
governed by the flow \eq{eq:PQM2+1flow}. The finite temperature correction can
then be obtained by integrating the vacuum flow from $\Lambda$ to
$\bar\Lambda\gg2\pi T$ and subsequently integrating the finite temperature
flow down to $k=\Lambda$. The resulting flow is especially easy for the
fermionic system, where one can actually use $\bar\Lambda=\infty$\,.
This procedure allows to study thermodynamic observables also at temperatures
above the phase transition. 
We refer the interested reader to \cite{Herbst:2013ufa} for a detailed
discussion.

\section{Results}
Although finite densities are straight-forward to implement in our setup, see
also \cite{Mintz:2012mz} for a discussion of possible issues, we focus in the
following on the case of vanishing chemical potential. On the one hand, this is
the region where we can benchmark our results with the lattice ones. On the
other hand, it also has numerical advantages, since the Polyakov loop and its
conjugate coincide in this case $\Phibar(T,\mu=0)=\Phi(T,\mu=0)$\,. This reduces
the numerical effort for solving the equations of motions drastically.

The effective potential in the infrared, $\Omega_{k\to0}$, evaluated at the
solution, $\chi_0=\left(\sigma_x, \sigma_y, \Phi\right)$, of the
corresponding equations of motion 
\begin{equation}
  \left.\frac{d\Omega_{k\to0}}{d\sigma_x}\right|_{\chi_0} =
  \left.\frac{d\Omega_{k\to0}}{d\sigma_y}\right|_{\chi_0} =
  \left.\frac{d\Omega_{k\to0}}{d\Phi}\right|_{\chi_0} = 0\,,
\end{equation}
will serve as the basis to calculate thermodynamic observables. It is related
to the pressure $P(T,\mu) = -(\Omega(T,\mu)-\Omega(0,0))$ with $\Omega(T,\mu) =
\Omega_{k\to0}(T,\mu)|_{\chi_0}$, and as such acts as a thermodynamic
potential, from which observables can be calculated in the standard manner.

To highlight the importance of fluctuations, we also show results from a
standard mean-field calculation, which neglects fluctuations of the mesonic
fields. It has been shown that the standard mean-field approximation (MF)
misses a contribution stemming from the fermionic vacuum loop. While this term is
naturally included in the FRG approach, it can be added to the standard MF,
resulting in the so-called extended MF (eMF) \cite{Skokov:2010sf,
Schaefer:2011ex}. 

Furthermore, it is our goal to check the proper inclusion of fluctuations as
well as unquenching effects. To do so, we compare our results to recent lattice
data by the HotQCD \cite{Bazavov:2011nk, Bazavov:2012bp} and
Wuppertal-Budapest collaborations \cite{Borsanyi:2010bp, Borsanyi:2010cj, 
Borsanyi:2013bia}.

In the following, we show all our results in terms of the reduced temperature
$t=(T-T_\chi)/T_\chi$, where $T_\chi$ denotes the chiral transition
temperature. This choice allows us to compare the overall shape - and
thereby the proper inclusion of the relevant dynamics - of the observables,
while a possible mismatch of the critical temperature is scaled out.
Note, however, that the scale mismatch is $\leq10\%$ within our FRG
calculation, see \tab{tab:Tc}.

\subsection{Phase Structure}
\begin{figure*}
  \includegraphics[width=.49\textwidth]{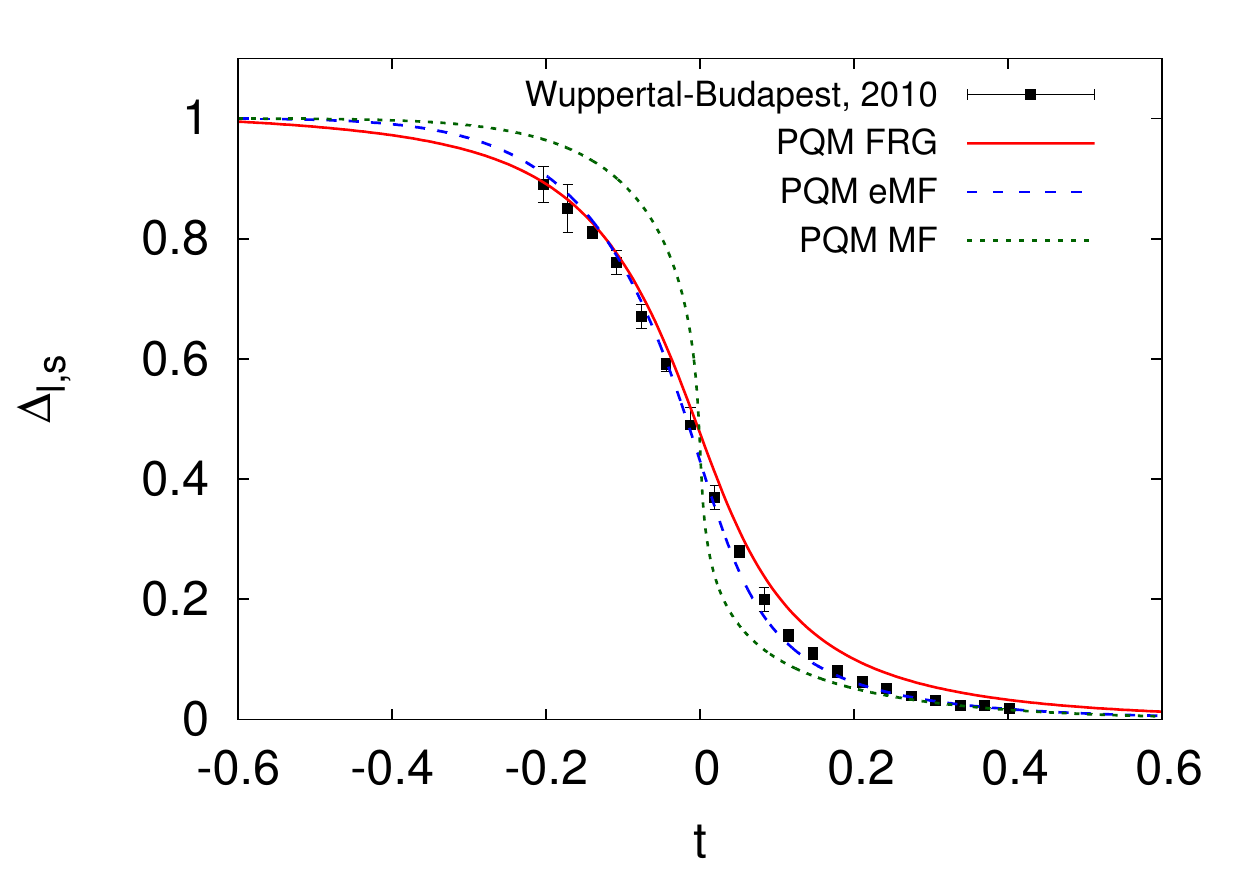}
  \includegraphics[width=.49\textwidth]{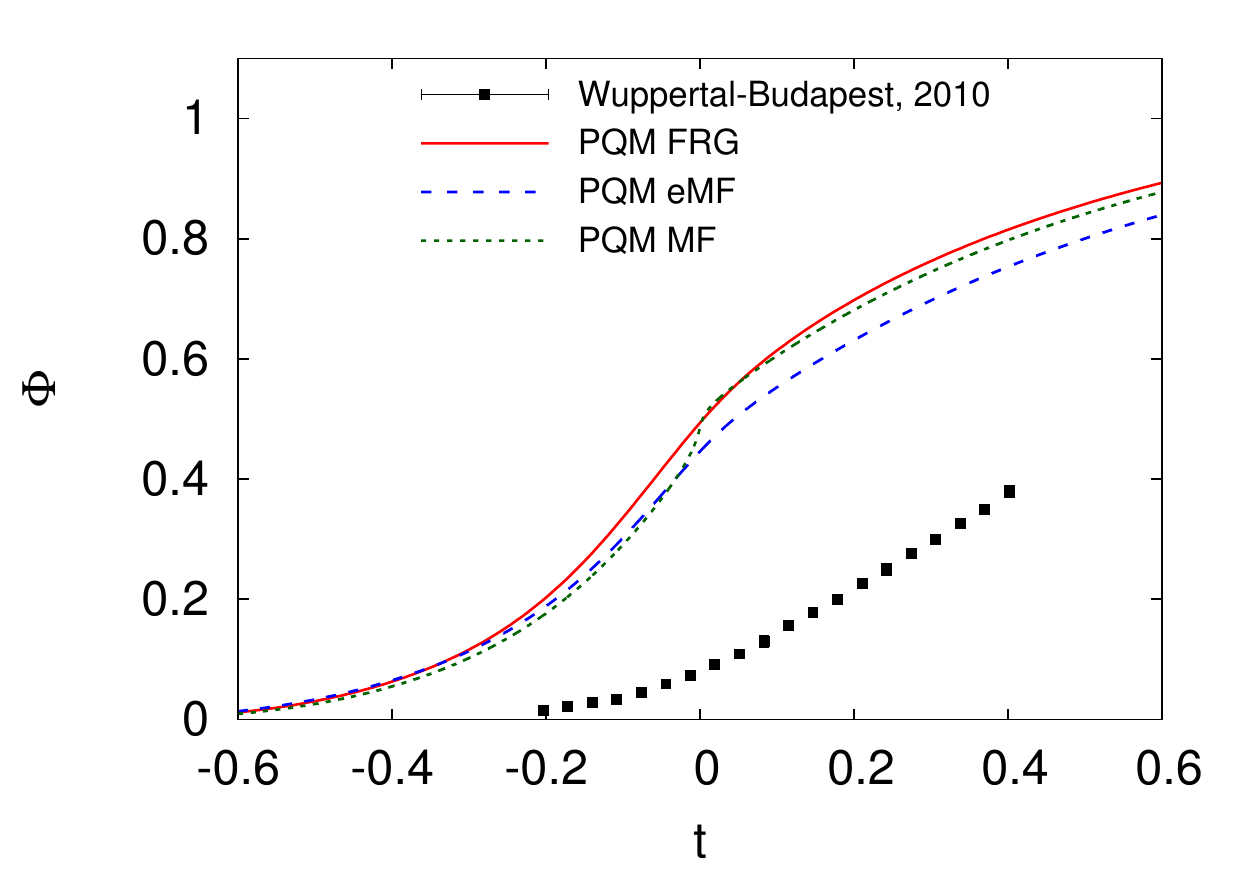}
  \caption{Temperature dependence of the subtracted chiral condensate (left)
  and  Polyakov loop (right). The FRG result is compared to the lattice one
  of the Wuppertal-Budapest collaboration, \cite{Borsanyi:2010bp}, as well as
  to mean-field computations. See text for details and comments on the 
  Polyakov loop in continuum approaches.}
  \label{fig:order_params}
\end{figure*}
First, we consider the phase structure, defined by the order parameters. In
particular, we display the subtracted chiral condensate
\begin{equation}
 \Delta_{l,s}  = \frac{\left(\sigma_x-\frac{c_{x}}{c_{y}}\sigma_y\right)_T}{
  \left(\sigma_x-\frac{c_{x}}{c_{y}} \sigma_y\right)_{T=0}}\,,
\end{equation}
as well as the Polyakov loop in \fig{fig:order_params}. The left panel shows
the subtracted chiral condensate in comparison to the lattice result by the
Wuppertal-Budapest collaboration \cite{Borsanyi:2010bp}. As expected, the MF
result (green, short-dashed curve) is too steep in the transition region. The
slope of the curve decreases when the vacuum term is included (eMF, blue,
long-dashed curve), and the result agrees nicely with the lattice points. Our
full FRG result (red, solid curve) has a similar slope and coincides very well
with the lattice at low temperatures. Above the phase transition, however, it
slightly overestimates the lattice values. We expect that this can be cured by
dynamical hadronisation \cite{Gies:2001nw, Gies:2002hq, Pawlowski:2005xe,
Floerchinger:2009uf}.

Considering the Polyakov-loop shown in the right panel of
\fig{fig:order_params}, we want to emphasise that the quantity considered on the
lattice, $\langle\Phi\rangle$, is different from the one calculated in our
continuum approach, $\Phi[\langle A_0\rangle]$. Based on the Jensen inequality
one can show that the two observables are related via $\Phi[\langle
A_0\rangle]\geq~\langle\Phi\rangle$ \cite{Braun:2007bx, Braun:2009gm,
Marhauser:2008fz}, i.e. the continuum observable serves as
an upper bound for the lattice one. Hence it is not surprising that all our
curves are far above the lattice points. Furthermore, we want to stress that the
confinement-deconfinement transition is a crossover in full QCD. Thus, there is
no unique definition of the critical temperature and different (approximate)
order parameters will likely show different transition temperatures. In
particular, the commonly used definition of the Polyakov loop on the lattice
contains a temperature-dependent normalisation, which can shift the crossover
temperature. This is not the case for the definition employed here.

For completeness we also quote our results, as well as the lattice ones, for
the critical temperature in \tab{tab:Tc}. We have used the peak in the
temperature derivative of the non-strange condensate/Polyakov loop as a
definition of the phase transition temperatures for the crossovers. Note that
the Polyakov loop on the lattice is rather flat, cf. \fig{fig:order_params} 
(right), and no deconfinement critical temperature is given.

\begin{table}
  \centering
  \begin{tabular}{c|c|c|c|c|c}
    Method & MF & eMF & FRG & WB \cite{Borsanyi:2010bp} & HotQCD 
      \cite{Bazavov:2011nk} \\\hline\hline
    $T_\chi$ [MeV] & 158 & 181& 172 & $157\pm{3}$ &$154\pm{9}$\\\hline
    $T_d$ [MeV] & 158 & 173& 163 & - &-
  \end{tabular}
  \caption{Chiral and deconfinement critical temperatures resulting from the
  different methods.}
  \label{tab:Tc}
\end{table}

\subsection{Thermodynamics}
Next, we discuss the pressure $P$ as well as the interaction measure $\Delta$,
which can be derived from the grand potential in the standard way:
\begin{equation}
  \Delta = \epsilon - 3 P\,,
\end{equation}
with $\epsilon = -P + T s + \sum_f \mu_f n_f \,,$
where $s=\partial P/\partial T$ and  $n_f = \partial P/\partial \mu_f\,, \ 
f=u,d,s $ are the entropy and quark number densities, respectively. As before,
we consider vanishing chemical potential, i.e. the third term in the energy
density is absent. 

We compare these quantities to results of the HotQCD collaboration,
\cite{Bazavov:2012bp} (open symbols), using the HISQ action and temporal lattice
extents of $N_\tau = 8, 12$ as well as to the continuum extrapolated results of
the Wuppertal-Budapest collaboration \cite{Borsanyi:2010cj} (full symbols).
Moreover, we also show the continuum fit provided by the Wuppertal-Budapest
collaboration \cite{Borsanyi:2013bia} (black, solid line).

\begin{figure*}
  \includegraphics[width=.49\textwidth] 
  {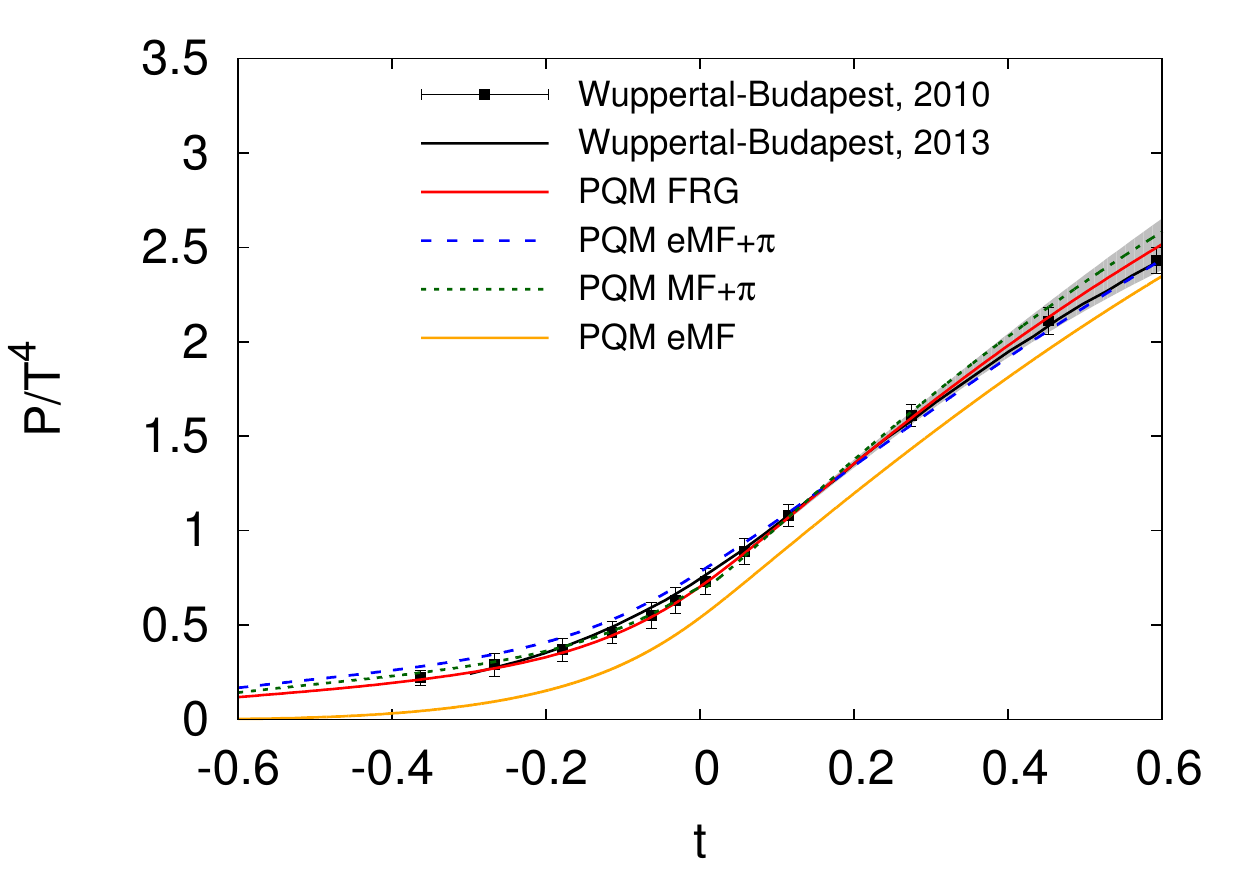}
  \includegraphics[width=.49\textwidth]
  {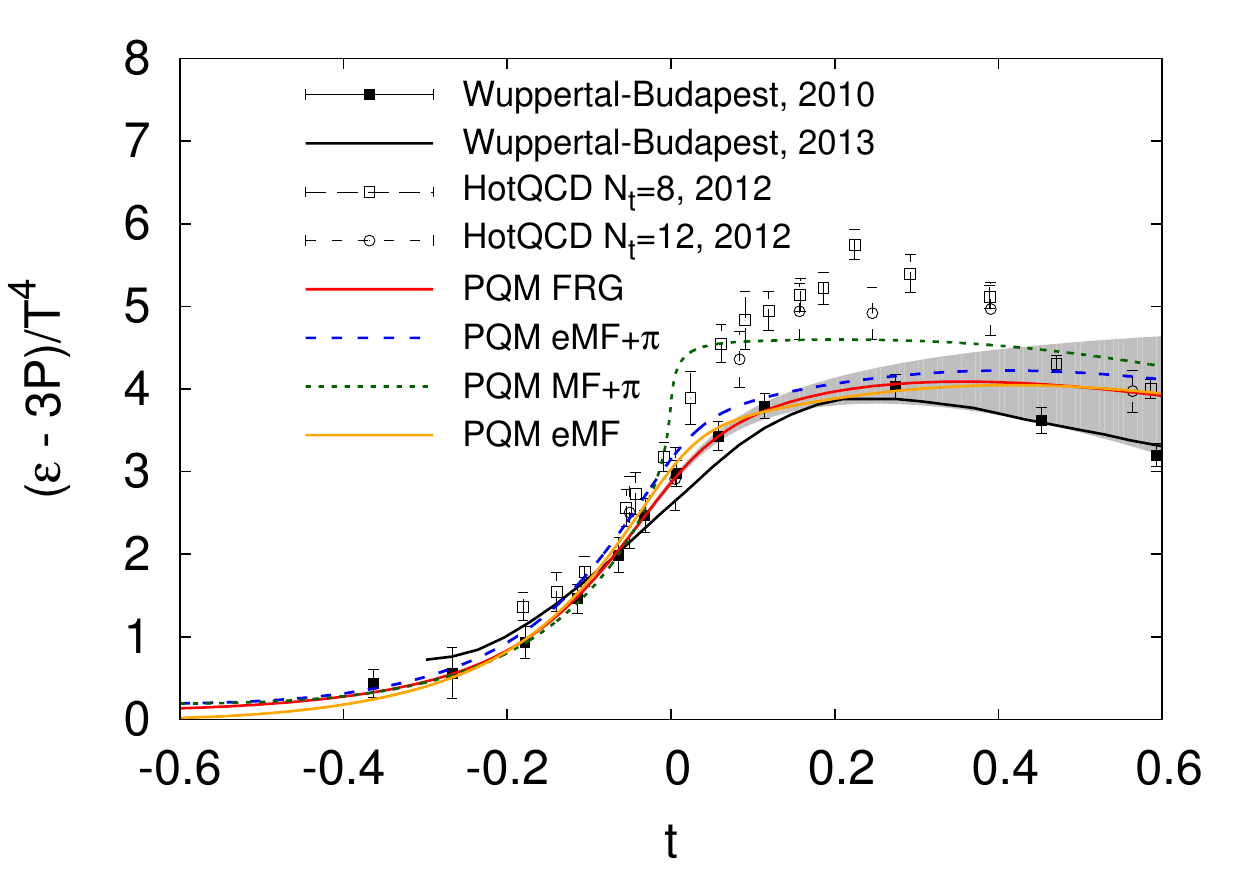}
  \caption{$(2+1)$-flavour FRG results for the pressure (left) and interaction
  measure (right) compared to the lattice, \cite{Bazavov:2012bp,
  Borsanyi:2010cj, Borsanyi:2013bia}, and  mean-field results. See text for
  details.}
  \label{fig:pdelta_poly}
\end{figure*}

Results for the pressure are displayed in \fig{fig:pdelta_poly} (left). As
before, we show results in mean-field approximations and by the FRG. First, we
discuss the eMF result depicted by the yellow, solid curve in
\fig{fig:pdelta_poly}. Since this approach neglects mesonic fluctuations, the
curve is approximately zero at small temperatures, where pions carry the
dominant contribution to the pressure. In order to overcome this deficiency, we
have chosen to supplement our MF and eMF computations with a gas of thermal
pions (MF+$\pi$: green, short-dashed; eMF+$\pi$: blue,
long-dashed). The pion in-medium mass is determined by the mean-field
potential. 
Strictly speaking, this induces a field-dependence in this
contribution, which would modify the equations of motion. Here, however, we
consider it as a correction to the thermodynamic potential only, and hence
neglect its backcoupling on the equations of motion.
For consistency, we also neglect all terms containing field derivatives,
$dP_\pi/d\phi_i$, in higher thermodynamic observables. 
In the FRG approach,
which includes mesonic fluctuations in a consistent manner, such a modification
is, of course, not necessary. The nice agreement of the mean-field with
our FRG result and the lattice at low temperatures confirms that it is indeed
crucial to include pionic modes.

As discussed above, we have used the pressure to fix the open parameter $T_{\rm
cr}^{\rm glue}$. This entails of course, that our results agree well with the
lattice results at $t=0$. Away from $t=0$, on the other hand, the agreement can
be attributed to the proper inclusion of the relevant dynamics. Especially
our FRG result (red, solid curve) agrees very well with the lattice throughout
the whole temperature range.
This can be seen more clearly in the interaction measure, shown in the right
panel of \fig{fig:pdelta_poly}. Clearly, the MF result is too steep in the
interaction region and its peak is too high as compared to the continuum
extrapolated lattice result by the Wuppertal-Budapest collaboration (full
symbols). The eMF result is still slightly to steep in the transition region.
Finally, our FRG result lies almost on top of the lattice points. 
This coincidence is not only due to the presence of matter fluctuations. In
fact, the inclusion of unquenching effects, $t_{\rm YM}(t_{\rm glue})$, results
in a drastic reduction of the peak height in the interaction measure towards the
lattice results, see also \cite{Haas:2013qwp, Stiele:2013gra} for a direct
comparison in MF.

At high temperatures, our results start to deviate from the lattice. However,
this is the region where the influence of the UV cutoff $\Lambda=1$~GeV sets in.
Furthermore, \eq{eq:tYMglue} ceases to be valid
at high scales where the slope saturates and the perturbative limit is reached.
The grey band in \fig{fig:pdelta_poly} gives an estimate of the error in our FRG
result. It is obtained from the change of the threshold functions with respect
to the temperature, at vanishing mass at the ultraviolet cutoff $\Lambda$, see
\cite{Herbst:2013ufa} for details.

\section{Conclusions}
In this talk we have discussed the inclusion of unquenching effects in the glue
potential in QCD effective models. Recent first-principles continuum results
for the unquenched glue potentials, \cite{Braun:2007bx, Braun:2009gm,
Fister:2013bh, Braun:2010cy}, have allowed to improve the glue sector of
effective models significantly, which had thus far been badly constrained:
mostly a Ginzburg--Landau-like ansatz for the glue potential, fitted to lattice
Yang-Mills theory, had been used. 
In \cite{Haas:2013qwp} it was shown that by a simple rescaling of the
temperature in the standard Yang-Mills based Polyakov-loop potentials one can
already capture the essential glue dynamics of the unquenched system. There,
results in the mean-field approximation have been put forward. Here, we have
taken the next step and included thermal and quantum fluctuations in the matter
sector with the functional renormalisation group. 

We have investigated order parameters and thermodynamic observables within the
Polyakov-extended quark-meson model with $2+1$ flavours. This type of
model can be systematically related to full QCD, as e.g.~discussed in
\cite{Haas:2013qwp, Herbst:2013ail, Pawlowski:2010ht}. A comparison to lattice
QCD simulations
with $2~+~1$~flavours shows excellent agreement up to temperatures of
approximately $1.3$ times the critical temperature. Therefore, we conclude that
most of the relevant dynamics for the QCD crossover can already be captured
within the PQM model.

The present work serves as a benchmark of our system at vanishing chemical
potential, which allows us to conclude that we have all relevant fluctuations
included. Since our approach is not restricted to the zero chemical potential
region we can now aim at the full phase diagram, $\mu\geq 0$\,.

\acknowledgments TKH is grateful to the organisers for making this
stimulating workshop possible.  The authors thank L.~Haas, L.~Fister
and J.~Schaffner-Bielich for discussions and collaboration on related
topics. This work is supported by the Helmholtz Alliance HA216/EMMI,
by ERC-AdG-290623, the FWF grant P24780-N27, by the GP-HIR, by the
BMBF grant OSPL2VHCTG and by the Helmholtz International Center for
FAIR within the LOEWE program of the State of Hesse.

\bibliographystyle{./utphys}
\bibliography{refs}

\end{document}